\documentclass[10pt]{article}

\usepackage{latexsym,cite,amsmath,epsfig,amsfonts}

\DeclareFontFamily{OT1}{rsfs10}{}
\DeclareFontShape{OT1}{rsfs10}{m}{n}{ <-> rsfs10 }{}
\DeclareMathAlphabet{\mathscript}{OT1}{rsfs10}{m}{n}

\numberwithin{equation}{section}

\newcommand{\be}{\begin{equation}}
\newcommand{\ee}{\end{equation}}

\newcommand{\bea}{\begin{eqnarray}}
\newcommand{\eea}{\end{eqnarray}}

\newcommand{\ns}{\normalsize}
\newcommand{\pt}{\partial}

\def\a{\alpha}
\def\b{\beta}

\def\f{\phi}

\def\l{\lambda}
\def\m{\mu}

\def\F{\Phi}

\def\L{\Lambda}

\begin{document}

%%%%%%%%%%%%%%%%%%%%%%%%%%%%%%%%%%%%%%%%%%%%%%%%%%%%%%%%%%%%%%%%%%%%%%
\begin{titlepage}

\vspace{-2cm}

\title{
\hfill{\ns OUTP-03-06P}\\
\hfill{\ns hep-th/0302024\\[.5cm]}
{\LARGE Inflation and K\"ahler Stabilization of the Dilaton}}

\author{{\ns\large David Skinner} \\[0.8em]
      {\ns Theoretical Physics, Department of Physics
      University of Oxford}\\[-0.2em]
      {\ns 1 Keble Road, Oxford OX1 3NP, United Kingdom} \\[-0.2em]
      {\it\ns skinner@thphys.ox.ac.uk}}
   
\date{}

\maketitle

\begin{abstract}

The problems of attempting inflationary model-building in a theory
containing a dilaton are explained. In particular, I study the shape
of the dilaton potential today and during inflation, based on a
weakly-coupled heterotic string model where corrections to the
K\"ahler potential are assumed to be responsible for dilaton
stabilization. Although no specific model-building is attempted, if
the inflationary energy density is related to the scale of gaugino
condensation, then the dilaton may be stabilized close enough to
today's value that there is no significant change in the GUT scale
coupling. This can occur in a very wide range of models, and helps to
provide some justification for the standard predictions of the
spectral index. I explain how this result can ultimately be traced to
the supersymmetry structure of the theory.

\thispagestyle{empty}

\end{abstract}
\end{titlepage}

%%%%%%%%%%%%%%%%%%%%%%%%%%%%%%%%%%%%%%%%%%%%%%%%%%%%%%%%%%%%%%%%%%%%%%%%

\section{Introduction}

The problem of a runaway dilaton in theories arising from the
low-energy limit of a string theory has been widely discussed. The
most frequently quoted mechanism for attempting to stabilize the
dilaton is via an effective potential generated when some hidden
sector gauge symmetry is driven to strong coupling. The gauginos
charged under the strong group condense~\cite{Ferrara:1982qs,Dine:rz},
with vevs that depend on the string-scale coupling and hence provide
an effective non-perturbative superpotential for the dilaton. If more
than one group has a gaugino condensate at comparable scales as in the
racetrack models~\cite{deCarlos:1992da,deCarlos:1992pd}, or if
non-perturbative string physics is accounted for as in K\"ahler
stabilization~\cite{Banks:1994sg,Binetruy:1996gg,Binetruy:1996nx,Casas:1996zi,Barreiro:1997rp},
then the dilaton may have a non-trivial minimum in its scalar
potential, as well as the supersymmetry preserving minimum at zero
coupling.

The question of why, during the cosmological evolution of our
universe, we should expect the dilaton to be found in the minimum at
non-trivial coupling is also well-known~\cite{Brustein:nk}. The
minimum generated by the above mechanism is typically rather shallow,
at least in comparison to the energies of the very early
universe. Generically one expects that, starting from an arbitrary
value, the dilaton would either not approach or else overshoot such a
minimum and roll away to zero coupling. Possible solutions to this
tend to invoke friction in the dilaton's equation of motion, generated
either from cosmological expansion~\cite{Barreiro:1998aj} or higher
Kaluza--Klein modes of the dilaton that are expected to be present at
high energies~\cite{Dine:2000ds}. Here I will concentrate on a
different issue: how do we know that, throughout the evolution of the
universe, the dilaton's potential had a minimum located in the same
place as today? Since each term in the effective supergravity scalar
potential couples to the dilaton, at energies in the early universe
above the current dilaton mass (typically of order 10-100 TeV), we may
expect that the dilaton potential has a different form. For example,
during inflation we expect that some term in this scalar potential
dominated the energy density of the universe. Since we are not
inflating today, there is no reason to expect that this same term is
still important and so it does not necessarily play any part in
dilaton stabilization today. How can we know whether such a term would
have provided a non-trivial dilaton minimum at all, let alone one
close to today's value?

This issue is important for various reasons. Of course, various
cosmological implications of varying coupling constants have been
widely discussed recently in {\it
e.g.}\cite{Webb:2000mn,Banks:2001qc,Bahcall:2003rh}, however here we
will consider issues relating more to inflation. Firstly, we do not
wish to exacerbate the Brustein--Steinhardt problem. For example, if
we make the very na\"{\i}ve approximation that the effective dilaton
potential instantaneously switched from its inflationary form to
today's form at the end of slow-roll inflation\footnote{In this paper
we make the assumption that the dilaton itself is not the inflaton.},
then we do not wish to have to guide it to its new minimum
again. Indeed, if this was originally done using higher KK modes as
in~\cite{Dine:2000ds}, then a reheat temperature high enough to
re-excite these would presumably also be high enough to generate all
manner of dangerous relics. Of course, this is just the usual
statement of the moduli problem~\cite{Banks:1993en} as applied to the
dilaton field. However it is also important for a second, possibly
more urgent reason specific to the dilaton (at least in the
weakly-coupled heterotic theory). Inflationary model building concerns
itself in part with generating sufficiently flat potentials for
slow-roll to occur. A promising way to achieve this without unnatural
tuning is to note that, in a globally supersymmetric theory, any
potential that is tuned to be flat at tree level will remain so to all
orders in perturbation theory so long as supersymmetry remains
unbroken. After SUSY breaking, such a potential will become sloped due
to quantum corrections. In SUGRA there is also the difficulty that
gravitational interactions generate a mass for canonical scalar fields
of the same order as their potential; the $\eta$ problem (see {\it
e.g.}~\cite{Lyth:1998xn}). Suggested ways of overcoming the
$\eta$-problem include the use of non-canonical forms for the
K\"{a}hler potential arising for fields well below the string
scale~\cite{Ross:1995dq,German:1999gi}, or identifying the slow-roll
and string moduli fields~\cite{Copeland:1994vg,Barreiro:1999hp}, or
perhaps by driving inflation via a
D-term~\cite{Binetruy:1996xj,Halyo:1996pp}. Assuming one or other of
these is successful, then the slope of the scalar potential will again
depend on quantum corrections and it is here that the dilaton plays a
crucial r\^ole. If these quantum corrections are to be under
reasonable calculational control then we had better ensure that during
inflation the dilaton is stabilized in the weak-coupling regime of
some theory. Even assuming that it is, clearly the spectral index of
primordial fluctuations $n_k$ will depend on the location of the
dilaton. This conclusion is true irrespective of any renormalization
of the coupling constants that must be accounted for in running down
from the string scale to the inflationary scale. Therefore, if we wish
inflationary model building to be at all predictive, it is essential
that we know where the dilaton is stabilized during inflation. Even
so, we may still expect that there will be some degeneracy in $n_k$
between different inflationary potentials and different dilaton
minima. With our present lack of knowledge about the form of the
potential during inflation, together with a far from completely
satisfactory explanation of how to stabilize the dilaton even today,
let alone during less well-understood epochs, it seems almost hopeless
to attempt to resolve this situation at present by anything other than
very model-dependent statements. Nonetheless, we will see that some
reasonably generic, natural mechanisms may provide significant help.

I will assume that the dilaton is stabilized via corrections to the
K\"ahler potential, perhaps arising through non-perturbative string
physics. If these are to be relevant in the field-theoretic regime,
they should be described in the K\"ahler potential by some function
$g(1/{\rm Re} S)$.  Much previous work in the subject has concerned
itself finding vacua that follow from a specific choice of this
function, such as those suggested by Shenker~\cite{Shenker}. However,
various forms are commonly discussed and there is even some debate as
to whether any of them realistically represent the true form of string
non-perturbative corrections. Here we will keep this function
completely arbitrary, except to assume that its presence in the
K\"ahler potential is capable of providing an acceptable minimum.  In
the next section, I briefly review the model of Bin\'etruy, Gaillard
and Wu~\cite{Binetruy:1996gg,Binetruy:1996nx} which treats K\"ahler
stabilization using in the linear multiplet formalism. Section 3
follows the standard path and considers the vacuum today. I will be
less interested here in particular examples, but rather will simply
list the usual constraints that the corrections to the K\"ahler
potential must satisfy if they are to be capable of providing a
phenomenologically reasonable vacuum, with {\it e.g.} vanishing
cosmological constant and SUSY breaking at $\sim$ 1 TeV. Having
considered the properties of the present vacuum, section 4 returns to
the full scalar potential and hence examines the vacuum during
inflation. Again I stress that we will not be concerned with building
specific inflationary models, but rather we will attempt to find a
reasonable generic scheme that ensures the dilaton potential has a
minimum during inflation. The key will be to assume that the scalar
fields acquire vevs which depend on the dilaton only via the
condensate itself. This is a natural mechanism, being the hidden
sector precursor to generation of visible sector SUSY breaking
masses. Additionally, in this picture the inflationary energy density
is always provided by the matter $F$-terms. This is a standard
assumption of much string-inspired inflationary model building, but
here it will follow from the above mechanism. Remarkably, the models
which have energy density compatible with the COBE bound $V^{1/4} \leq
6.7 \epsilon^{1.4} \times 10^{16}$ GeV also tend to stabilize the
dilaton close to today's value. This is true despite the many orders
of magnitude difference between the effective cosmological constants
during inflation and today. Similar ideas were pursued
in~\cite{Gaillard:1998xx} where the authors allowed scalar field vevs
to be generated by vacuum shifting to cancel a D-term. However, the
results here are more general, and in particular do not depend on any
form for the K\"ahler corrections. Finally, the conclusions will
discuss the obtained results and try to understand why they hold. I
argue that ultimately, this can be traced to the supersymmetry
structure of the higher dimensional theory.

%%%%%%%%%%%%%%%%%%%%%%%%%%%%%%%%%%%%%%%%%%%%%%%%%%%%%%%%%%%%%%%%%%%%%%%%%

\section{The model}

In this paper I assume that, both today and during inflation, the
dilaton is stabilized through corrections to the K\"ahler potential.
For convenience, we will work in the linear supermultiplet formalism
of~\cite{Binetruy:1996gg,Binetruy:1996nx},  although since this has been
shown to be equivalent to the usual chiral formulation in
ref~\cite{Burgess:1995kp}, we expect the same results  to hold there
also. In this formalism, for each semi-simple hidden sector gauge
group ${\mathcal G}_a$ the gaugino condensate superfields $U_a \sim
{\rm Tr}(W^\a W_\a)_a$ are identified with the chiral projections of
the vector superfields as
\bea
U_a &=& -({\mathcal D}_{\dot{\a}}{\mathcal D}^{\dot{\a}} - 8R) V_a 
\nonumber \\
\overline{U}_a &=& -({\mathcal D}^{\a}{\mathcal D}_{\a} - 8\overline{R}) V_a 
\eea
with the overall vector superfield $V = \sum_a V_a$ having the dilaton
$l$ as its lowest component: $l = V|_{\theta = \bar{\theta} = 0}$. Compared 
to the chiral formulation, we have roughly $l \approx 1/(S+\overline{S})$.
Considering for simplicity an orbifold compactification, the K\"ahler 
potential is
\be
K = \ln(V) + g(V) - \sum_I \; \ln(T_I + \overline{T}_I - \sum_A |\F_{AI}|^2) 
\label{Kahler}
\ee
where here I consider only untwisted chiral matter fields $\F_{AI}$. The
function $g(V)$ is supposed to describe the contribution of string
non-perturbative effects to the K\"{a}hler potential, and clearly this
affects also the dilaton kinetic term. A simple calculation shows this
to be
\be
{\mathcal L}_{\rm kinetic} \supset \frac{(lg^{\prime}+1)}{4l^2}
\pt_\m l \, \pt^\m l
\label{dilkin}
\ee
where the prime denotes differentiation with respect to $l$.  The
string non-perturbative effects also modify the string-scale gauge
coupling of the effective  field theory:
\be
2\pi\a = \frac{l}{1+f(l)}
\label{coupling}
\ee
and in the 4D Einstein frame, $f$ is related to $g$ via
\be
V\frac{dg}{dV} = f(V) - V\frac{df}{dV} \; ,
\label{Einst}
\ee
with $f(0)=g(0)=0$ as required in the weak-coupling limit.

The gaugino condensates are described by the Veneziano-Yankielowicz
superpotential~\cite{Veneziano:1982ah,Ferrara:1990ei}, appropriately
generalised to incorporate SUGRA as well as the presence of several
gaugino condensates and/or gauge-invariant matter condensates. There
are  also superpotentials for the matter fields and terms describing
quantum corrections to any possible unconfined gauge groups. It is of
course important to preserve the modular invariance of the underlying
string theory, and a Green-Schwarz counterterm is introduced to this
end (it is assumed that the modular anomaly is completely cancelled by
this term). The reader is referred to
refs.~\cite{Binetruy:1996gg,Binetruy:1996nx} for details.

Upon solving the equations of motion, the authors find the following
expression for the condensates $u_a = U_a|_{\theta = \bar{\theta} = 0}$
\be
|u_a|^2 = A^2\exp \left[\ln(l) + g(l) - \frac{(1+f)}{b_a l}
+ \frac{b_{E_8}-b_a}{b_a} \sum_I \ln x_I \right]
\prod_I 
\left| \eta(t_I) \right|^{4(b_{E_8}-b_a)/b_a}
\label{cond}
\ee
where $t_I = T_I|_{\theta=\bar{\theta}=0}$, $x_I \equiv t_I +
\overline{t}_I - \sum_{A} |\f_{AI}|^2$ and $\eta(t_I)$ is the Dedekind
$\eta$-function, which ensures that \eqref{cond} is modular invariant.
The constants $b_a$ and $b_a^\prime$ are related to the $\b$-functions
for the condensing gauge group ${\mathcal G}_a$. In
particular~\cite{Binetruy:1996nx},
\be
b_a = \frac{1}{8\pi^2} \left( C_a - \frac{1}{3}\sum_A C^A_a\right)
\label{b}
\ee
where $C_a$ and $C^A_a$ are the quadratic Casimirs for the group $G_a$
in the adjoint and matter represenations, respectively. This is
$1/8\pi^2$ times the co-efficient of the one-loop $\b$-function for
the coupling, and gives $b_{E_8} = 30/8\pi^2 \approx 0.38$ whereas,
for typical choices of hidden sector groups and matter
representations~\cite{Gaillard:1999et},  $b_a \leq 0.1$. The constant
of proportionality $A^2$ in \eqref{cond} again depends on the group
structure, matter representations and Yukawa couplings present in the
entire model; we allow this the fairly generous range $10^{-4} - 10^4$
(see~\cite{Gaillard:1999et}) though its value will not be crucial
here. Notice that the condensate depends on the dilaton explicitly
through the exponentials of both the K\"ahler potential and the
inverse of the string-scale gauge coupling, and implicitly through any
dilaton dependence of the matter fields $\f_{AI} =
\left.\F_{AI}\right|_{\theta = \overline{\theta} = 0}$ or $t_I$
moduli. This of course follows from the interpretation of the
condensate scale as the energy at which the logarithmic running of
$\a$ causes it to become strong. As pointed out in~\cite{Dine:1999dx}
for the whole perturbative model to make sense in the first place, we
of course require that the theory started out from reasonably weak
coupling, and so we assume without further justification $b_a l \leq 1$ at 
all times.

One practical advantage of K\"ahler stabilization is that, even in the
presence of several condensing gauge groups, the physical properties
of the vacuum are dominated by the condensate coming from ${\mathcal
G}_+$ the group with the largest beta function coefficient $b_+$. 
The condensate scale $\L_c$ and gravitino mass $\tilde{m}$ are then 
given by
\bea
\L_c &=& \langle |u_+|^2 \rangle^{\frac{1}{6}} \\
\tilde{m} &=& \frac{b_+}{4} \langle |u_+|\rangle
\label{scales}
\eea
in reduced Planck units. So, as usual, the condensate plays the
r\^{o}le of an effective non-perturbative superpotential for the
dilaton $b_+ u_+/4 \sim e^{K/2} W_{\rm np}$.

The final expression we require is the effective scalar potential for
the model. This was presented in~\cite{Gaillard:1998xx}. Neglecting
the twisted sector matter fields, it is given by the somewhat complicated
form
\begin{multline}
V = \frac{e^K}{1+b_{E_8}l} \sum_I \; \left[ \left|(2\xi(t_I) x_I +1)B_I 
-\sum_A \f_{AI}\frac{\pt W}{\pt \f_{AI}} \right|^2 + 
x_I \sum_A \; \left| \frac{\pt W}{\pt \f_{AI}} +
2\xi(t_I) B_I \overline{\f}_{AI} \right|^2 \right] \\
+ (lg^\prime +1) e^K \left|
\frac{b_+u_+e^{-K/2}}{4}\left(1+\frac{1}{b_+l}\right)- W \right|^2
-3 e^K \left| \frac{b_+ u_+e^{-K/2}}{4} - W \right|^2
\label{V}
\end{multline}
where $\xi(t_I) = \frac{1}{\eta(t_I)}\frac{d\eta}{dt_I}$ and $B_I$ 
is defined by
\be
B_I = \sum_A \f_{AI}\frac{\pt W}{\pt \f_{AI}} - W - e^{-K/2}\frac{u_+}{4}
(b_{E_8}-b_+) \; .
\label{Bdef}
\ee
Modular invariance of the scalar potential requires that, up to a 
possible modular invariant function that would contain singularities, 
the superpotential has the form
\be
W = \sum_n c_n \prod_{AI}
\f_{AI}^{p_n^{AI}}\eta(t_I)^{2(p_n^{AI}-1)}
\label{W}
\ee
for some powers $p_n^{AI}$ with Yukawa couplings $c_n$ that we will
assume are ${\mathcal O}(1)$. For a cubic superpotential, we take
$\sum_{AI} p_n^{AI} = 3$ in each term $n$.

In comparison with the usual chiral formulation, the first line of
\eqref{V} approximately corresponds to the $F$-terms for the $t_I$-moduli
and matter fields, while the second line contains the dilaton $F$-term
together with the usual $-3 e^K |W|^2$ piece, bearing in mind the
interpretation of the condensate as a non-perturbative contribution
to the superpotential. Notice also the factor involving
$1/(1+b_{E_8}l)$  which multiplies the moduli and matter $F$-terms;
this arises from the Green-Schwarz counterterm preserving modular
invariance in the theory.

%%%%%%%%%%%%%%%%%%%%%%%%%%%%%%%%%%%%%%%%%%%%%%%%%%%%%%%%%%%%%%%%%%%%%%%%%

\section{The dilaton potential today}

Following the standard assumptions, we assume that at least one matter
field in each term in the superpotential and its derivatives has zero
vev today (or at least has a vev $\ll \L_c$). Minimizing the remaining
potential with respect to $t_I$, the moduli are
found~\cite{Binetruy:1996nx} to be located at their self-dual points
where $(4\xi \, {\rm Re} (t_I)+1)=0$. This is in accordance with the
general result which states that these self-dual points are always
extrema of the scalar potential~\cite{Shapere:1988zv,Font:1990nt}. Such a
stabilization of the $t_I$-moduli causes their $F$-terms to vanish and
accordingly in these models SUSY is broken by a non-vanishing dilaton
$F$-term. Thus the remaining scalar potential is
\be
V = \left[(lg^\prime +1)\left(1+\frac{1}{b_+l}\right)^2 - 3\right]
\frac{b_+^2|u_+|^2}{16} \; .
\label{Vtoday}
\ee

In order to obtain a phenomenologically viable model, we would now wish 
to impose that the dilaton is at the minimum of its potential, in a
vacuum with zero cosmological constant and an acceptable values for the
gauge coupling and gravitino mass. This corresponds to the conditions
\bea
\left.(1+b_+l)^2(lg^\prime+1) - 3b_+^2 l^2\right|_{l=l_0} &=& 0 
\label{V=0}\\
\left.f^{\prime\prime}+\frac{6b_+^2}{(1+b_+l)^3}\right|_{l=l_0} &=& 0
\label{Vextr}\\
\left.f^{\prime\prime\prime} - \frac{18b_+^3}{(1+b_+l)^4}\right|_{l=l_0}
 &<& 0 \label{Vmin}
\eea
where \eqref{V=0} corresponds to the vanishing of the remaining scalar 
potential and \eqref{Vextr}-\eqref{Vmin} ensure that the dilaton is 
at a minimum. Additionally, constraints on the GUT scale coupling and SUSY
breaking scales may be interpreted as requiring
\bea
\left.\frac{l}{2 \pi (1+f)}\right|_{l=l_0} = &\a_0& = \frac{1}{25}
\label{alpha}\\
\L_c = 10^{13}\; {\rm GeV} \; &\Rightarrow& \; \tilde{m} = 1\; 
{\rm TeV}\; .
\label{mgrav}
\eea
As mentioned in the introduction, much previous work on these models
has now specified particular forms ({\it e.g.} those of
\cite{Shenker}) for the function $f(l)$ (and hence also $g(l)$) so as
to proceed to find viable examples. While it is certainly interesting
that this can be done in practice, here I shall pursue a different
route and examine the implications of the above constraints more
generally. 

Firstly, notice that \eqref{V=0} implies
\be
lg^{\prime} + 1 > \frac{3b_+^2l^2}{(1+b_+l)^2}
\label{constraint}
\ee
for $l\neq l_0,\, 0$ as otherwise there would be a vacuum with lower 
energy\footnote{Of course we also have $V=0$ at $l=0$; the SUSY 
preserving, uncoupled vacuum.}. Next consider the gauge coupling. 
Expanding $f(l)$ around $f(l_0)$ gives
\be
\frac{1}{2\pi \a} = \frac{1}{2\pi \a_0} - 
\frac{3b_+^2}{(1+b_+l_0)^2}(l-l_0) +\frac{1}{l_0}\sum_{n=3}^{\infty}
\left.\frac{d^nf}{dl^n}\right|_{l=l_0} \frac{(l-l_0)^n}{n !}\; ,
\label{da}
\ee
where the first and second order terms are evaluated explicitly using
\eqref{V=0}-\eqref{Vextr}. Eq \eqref{da} determines how the GUT scale
coupling changes if the dilaton is located away from its true
minimum. What is important to notice is that the first and
second order terms always induce a negligible change in $\a$ for
$b_+|l-l_0| \ll 1$, in other words throughout the entire
``weakly-coupled'' regime. Any significant change in the GUT scale
coupling is therefore traceable to the presence of large higher order
derivatives in the expansion \eqref{da}. Indeed, such terms can
generically be present, since eqs~\eqref{V=0}-\eqref{Vmin} do not
restrict their magnitude. To paraphrase: if $\a$ does not change
significantly during inflation, then $|l_i -l_0|$ is sufficiently
small that higher order derivatives of $f$ are unimportant. We shall
see if this is reasonable in the next section.

%%%%%%%%%%%%%%%%%%%%%%%%%%%%%%%%%%%%%%%%%%%%%%%%%%%%%%%%%%%%%%%%%%%%%%%%%

\section{The dilaton potential during inflation}

Let us now proceed to examine the dilaton potential during inflation.
Since we do not know {\it a priori} which term is responsible for 
inflation, at first sight this seems to require the somewhat daunting
task of investigating the full scalar potential~\eqref{V}. However the 
COBE bound requires
\be
V^{1/4} \leq 6.7\epsilon^{1/4} \times 10^{16} \; {\rm GeV}\;,
\label{COBE}
\ee
with typical slow-roll parameters $\epsilon$ giving $V^{1/4} \approx
10^{14}$ GeV. If we assume this energy density is provided by the vevs
of some scalar field(s) as in hybrid inflation (see {\it e.g.}
\cite{Copeland:1994vg}), then of course we require $\langle
\f_{AI}\rangle \ll 1$ and therefore $W \ll \pt W /\pt \f_{AI}$ for
${\mathcal O}$(1) Yukawa couplings. This simplifies the potential to
\begin{multline}
V = \frac{e^K}{1+b_{E_8}l} \sum_I \; \left[ \left|(2\xi x_I +1)
e^{-K/2}\frac{u_+}{4}(b_{E_8}-b_+)\right|^2
+ x_I \sum_A \; \left| \frac{\pt W}{\pt \f_{AI}}\right|^2 \right] \\
+ (lg^{\prime} +1) e^K \left|
\frac{u_+b_+e^{-K/2}}{4}\left(1+\frac{1}{b_+l}\right) \right|^2
-3 e^K \left| \frac{b_+ u_+e^{-K/2}}{4} \right|^2 \; .
\label{Vinfl}
\end{multline}

For the theory to be under control, we must also have $|u_+| \ll 1$ in
reduced Planck units, but notice that we cannot simply neglect the
condensate; until we know where the dilaton is stabilized we do not
know the condensate scale in comparison to $\frac{\pt W}{\pt \f_{AI}}$
(of course, as yet there are no observational constraints on $\L_c$
during inflation). To proceed further, in this paper I will make the
fairly natural assumption that a typical scalar field acquires a vev
connected to the condensate scale $\L_c$, at whatever value this turns
out to be.\footnote{In~\cite{Gaillard:1998xx,Cai:1999aj} the authors
followed an alternative route and obtained the inflationary scale by
vacuum shifting to cancel a $D$-term.} A natural expectation would be
$\langle \f_{AI} \rangle \propto \L_c$ for some scalar field(s)
$\f_{AI}$. However, there may be some fields which acquire vevs
proportional to some other power of the condensate scale, perhaps due
to a global symmetry which restricts the way they appear in the
superpotential (see {\it e.g.}  \cite{Ross:2002mr}). For this reason, let 
us take a typical scalar vev to be
\be
\langle |\f_{AI}|^2 \rangle \propto |u_+|^{2\l} |\eta(t_I)|^{-4}
\label{vev}
\ee
where $\l$ is any positive power and the Dedekind $\eta$-function is
present to ensure the correct behaviour under modular transformations
(the condensate itself being invariant). Of course, this is exactly
the mechanism that generically provides SUSY breaking masses in the
observable sector today (with $\L_c \rightarrow \tilde{m}$ via gravity
mediation). Additionally, during inflation this mechanism has the
advantage of being able to provide a high inflationary energy scale in
a consistent way. The point is that in a string theory, all mass
scales under $m_s$ (in particular the inflationary scale $V^{1/4}$)
must be generated dynamically. A condensate or gauge-symmetry breaking
scalar field vev will only form if it is energetically permitted. This
may be inhibited in the presence of a large inflationary potential
from some other source, and so inflation would be in danger of
preventing its own SUSY breaking cause~\cite{Ross:1995dq}. Of course,
it is not necessary for all the fields to acquire the same vevs with
the same powers $\l$; this will be discussed further later. At
present, let us assume that all sums are restricted to their dominant
term(s). Again, I have ignored the possibility of additional modular
invariant functions in the vevs, and $\mathcal O$(few) numerical
factors will not affect the argument.

At first sight, it appears that there are now various cases to
consider; depending on $\l$, with $|u_+|^2\ll 1$ we may consider the
dominant contribution to $V$ to originate from either the matter,
moduli or dilaton $F$-terms during inflation. However, let us consider
the stabilization of the $t_I$-moduli. With the scalar vevs as in
\eqref{vev} and using \eqref{cond} we have
\bea
\frac{\pt |u_+|^2}{\pt t_J} &=& \frac{b_{E_8} - b_+}{b_+}
\left[\sum_I \frac{1}{x_I}\frac{\pt x_I}{\pt t_J} + 2 \xi(t_J) \right]
|u_+|^2\\
&=& \frac{b_{E_8} - b_+}{b_+}
\left[\frac{ 2\xi(t_J) (t_J+\overline{t}_J)+1}{x_J} |u_+|^2 -\l
\frac{\pt |u_+|^2}{\pt t_J} \sum_{AI}
\frac{\langle |\f_{AI}|^2\rangle}{x_I} \right]
\label{tderiv1}
\eea
and so the condensate is extremised at the self dual points where
$4\xi(t_J) \; {\rm Re} t_J + 1 = 0$. Now, from \eqref{Vinfl} it is
clear that these points also extremise the dilaton $F$-term and the
supergravity $-3|W_{\rm np}|^2$ term, as these only depend on the
moduli via the condensate. Consider next the matter $F$-terms. With a
cubic superpotential of the form \eqref{W} and matter vevs \eqref{vev}
we have
\be
\left| \frac{\pt W}{\pt \f_{AI}}\right|^2 = |u_+|^{4\l}|A_{AI}|^2
\prod_{J \neq I}\left|\eta(t_J)\right|^{-4}
\label{dW}
\ee
where $|A_{AI}|^2 = \left|\sum_{n} c_n p_n^{AI}\right|^2$ is a constant.
Hence for the matter $F$-terms we find
\bea
e^{-K}(1+b_{E_8}l)\frac{\pt V}{\pt t_J} &\supset& \sum_{A,I}\left[
\frac{\pt K}{\pt t_J}x_I + \frac{\pt x_I}{\pt t_J} +
x_I\frac{\pt \;}{\pt t_J}\right] \left|\frac{\pt W}{\pt \f_{AI}}\right|^2
\nonumber \\
&=& \sum_{A,I} x_I \left|\frac{\pt W}{\pt \f_{AI}}\right|^2 \left[
\sum_K -\frac{1}{x_K}\frac{\pt x_K}{\pt t_J} + \frac{1}{x_I}
\frac{\pt x_I}{\pt t_J} + 2\xi(t_J) (\delta_I^J -1) \right] + \ldots
\nonumber \\
&=& \sum_{A, I \neq J} -\frac{x_I}{x_J}
\left|\frac{\pt W}{\pt \f_{AI}}\right|^2 
\left[ 2\xi(t_J) (t_J+\overline{t}_J) + 1 \right] + \ldots
\label{Tstab}
\eea
where the ellipses represent terms proportional to $\pt |u_+|^2 /
\pt t_J$. Hence the matter $F$-terms are also minimized at the
self-dual points. Considering finally the moduli $F$-terms, from
\eqref{Vinfl} it is clear that these are {\it not} minimized at the
self-dual points, but rather where $2\xi(t_I)x_I + 1 = 0$. However,
at the self-dual points the only remaining pieces of the moduli
$F$-terms are
\be
\frac{(b_{E_8}-b_+)^2}{4(1+b_{E_8}l)} \sum_{I} 
\left|\xi(t_I) u_+ \sum_A|\f_{AI}|^2 \right|^2 \; .
\label{Fmod}
\ee
Since $|u_+|^2 \ll 1$ for the theory to be under control, it is clear
that \eqref{Fmod} is negligible compared to either the dilaton or
matter $F$-terms in \eqref{Vinfl} for generic scalar field vevs of the
form \eqref{vev}, irrespective of $\l$. Hence the overall potential is
minimized\footnote{It may further be shown that the points $t_I =
e^{i\pi/6}$ are minima, whereas $t_I = 1$ are saddle points of the
potential; see~\cite{Shapere:1988zv,Font:1990nt}.} when $4\xi(t_I){\rm
Re}(t_I) + 1=0$ and we may neglect the moduli $F$-terms
henceforth. This approximation will of course eventually cease to be
valid as the universe evolves - the remaining small pieces from the
moduli $F$-terms will eventually destabilize the potential and drive
$\langle |\f_{AI}|^2\rangle \rightarrow 0$ for at least one field in
each term in the superpotential and its derivatives, as is the case in
the true vacuum today.

Let us now return to our primary goal - the stabilization of the
dilaton during inflation. We have seen that, whatever the value of
$\l$ the inflationary potential will be dominated by either the matter
or dilaton $F$-terms. With $t_I = e^{i \pi/6}$ these are
\bea
V &=& \frac{e^K}{1+b_{E_8}l} |u_+|^{4\l}|\eta(e^{i\pi/6})|^{-8}
\sum_{A,I} |A_{AI}|^2 + V_0 \nonumber \\
&=& C\frac{le^g}{1+b_{E_8}l}|u_+|^{4\l} + V_0
\label{Vinfl2}
\eea
for $C = |\eta(e^{i\pi/6})|^{-8}\sum_{A,I} |A_{AI}|^2 /3$ where with
$\mathcal O$(1) Yukawa couplings we expect $1 \leq C \leq 10$. In
\eqref{Vinfl2}, $V_0$ is the form of the scalar potential today as
given by \eqref{Vtoday}, although of course as yet we do not know its
value during inflation.

As explained in the introduction, if one is to justify the standard
predictions of inflationary models whose slow-roll parameters come
from higher-order, coupling-constant dependent terms in the potential
(perhaps arising as quantum corrections) then it is necessary that the
dilaton be stabilized somewhere near to its value today. Here we do
not attempt solve the Brustein-Steinhardt problem~\cite{Brustein:nk}
and explain how the dilaton dynamically settled into this minimum
({\it e.g.} rather than the one at $l=0$ which is always present), but
simply try to ensure that if it {\it can} be solved before inflation,
it will not arise again afterwards. In order to achieve a minimum in
the dilaton potential close to the one today, it again appears that
there are two separate cases to consider. Firstly, $\l$ may be large
enough that the matter $F$-terms are typically much smaller than the
$V_0$ contribution (at values of $l \neq l_0$), and therefore $V_0$
would then drive $l\rightarrow l_0$ irrespective of the matter
term. Since we must have $|u_+| \ll 1$ for a sub-Planckian condensate,
this will occur when $\l > 1/2$ as then the matter $F$-terms are
typically suppressed compared to the $V_0$ term. Secondly, it may be
that the matter $F$-term itself has a minimum close to $l=l_0$. To
examine this case, consider the matter $F$-terms. Using eqs
\eqref{Einst} and \eqref{cond}, these are minimized when
\bea
\left. lg^{\prime} + 1 \right|_{l=l_i} &=& \left. 
\frac{b_{E_8} b_+ l^2}{(2\l(1+b_+l)+b_+l)(1+b_{E_8}l)}\right|_{l=l_i}
\nonumber \\
&\approx& \left.\frac{b_{E_8}b_+l^2}{2\l + b_+l}\right|_{l=l_i}
\label{dilmin}\; ,
\eea
where in the second line I make the usual weak-coupling approximations
$b_{E_8}l_i, \, b_+l_i \ll 1$. Comparison with \eqref{constraint} shows 
that such a value can only exist if
\be
\l < \frac{b_{E_8}}{6b_+}\; .
\label{lambda}
\ee 
The point is that the functions $f(l)$ and $g(l)$ are supposed to be
determined by non-perturbative aspects of the underlying string
model. We have specified phenomenologically desirable properties in
section 3, but once these are fixed, we are not free to retune the
functions during inflation\footnote{In particular this is true if the
$t_I$ moduli are fixed at the same place during inflation as
today, even if $g(l) \rightarrow g(l,t_I)$, though it is not clear that 
this would allow such a stabilization.}. Expanding $lg^{\prime}$ 
around $l=l_0$ gives
\be
\frac{b_{E_8}b_+l_i^2}{2\l + b_+l_i} = 3b_+^2l_0^2 + 6b_+^2l_0(l_i-l_0)
+ \sum_{n=2}^{\infty} \left.\frac{d^n (lg^\prime)}{dl^n}\right|_{l_0} 
\frac{(l_i-l_0)^n}{n!}
\label{lgprime}
\ee
where I have used eqs.~\eqref{V=0}-\eqref{Vextr} and \eqref{dilmin} to
evaluate the zeroth and first order terms. In this equation, the
derivatives in the sum are (linear combinations of) those in
\eqref{da}, evaluated at the same place $l=l_0$ and are therefore of
the same magnitude. Since the explicit terms on the {\it lhs} and {\it
rhs} of \eqref{lgprime} are much smaller than those of \eqref{da}, it
is clear that the difference $|l_i - l_0|$ between the dilaton minima
during inflation and today will not cause a significant change in
$\a$. Additionally, for typical
compactifications~\cite{Gaillard:1999et} the {\it rhs} of
\eqref{lambda} is at least 2/3 so this case smoothly overlaps with the
previous one where $l \rightarrow l_0$ because of the presence of
$V_0$. We therefore have the remarkable conclusion that {\it all} vevs
of the form \eqref{vev} will stabilize the dilaton at a value with $\a
\approx \a_0$. As explained in the introduction, this provides some
justification for the usual predictions of inflationary
model-building.

Let us now investigate the the dilaton mass. In comparing this to the
Hubble rate, it is of course important to consider the field $D$ with
canonical kinetic terms. Eq. \eqref{dilkin} shows that this is given by
\be
\frac{1}{4}\pt_\m D\, \pt^\m D = \frac{(lg^{\prime}+1)}{4l^2}
\pt_\m l \, \pt^\m l
\label{canon}
\ee
and hence we now define
\be
\eta_D \equiv \left.\frac{1}{V}\frac{d^2 V}{dD^2}\right|_{l_i} = 
\left.\frac{1}{V}\left(\frac{dl}{dD}\right)^2 
\frac{d^2 V}{dl^2}\right|_{l_i} \; .
\label{mass}
\ee
Again using the approximation $l_i=l_0$ and now considering the case
where $\l < 1/2$ so that the matter term dominates in its own right,
we have
\be
\eta_D \approx (lg^{\prime})^{\prime} 
\; \frac{(2\l+b_+l_i)^2}{b_{E_8}b_+^2l_i^2}
-\frac{4\l +1}{b_+l_i} 
\label{mass2}
\ee
and so in order to hold the dilaton in place at its minimum during 
inflation we require
\be
\left.(lg^{\prime})^{\prime}\right|_{l_i} \gg 
\frac{4\l+b_+l_i}{(2\l+b_+l_i)^2}\, b_{E_8}b_+l_i
\label{mass3}
\ee
which is not unreasonable, given that $b_+l_i \ll 1$. By comparison,
at $l=l_0$ we have $(lg^{\prime})^{\prime} \approx 6b_+^2l_0$ but we
may expect it to be significantly larger here as higher order curvature
terms come into play.

Finally, consider the inflationary energy scale. The energy
density during inflation will always be of the order of the matter
$F$-terms. This is because either this term dominates in its own right
(for $\l < 1/2$) or $V_0$ dominates and attempts to minimize itself at
$l=l_0$ where $V_0 = 0$, with this minimization now leaving a
residual piece of the order of the matter $F$-terms. This is in 
accordance with the standard assumptions of inflationary
model-building~\cite{Copeland:1994vg,Lyth:1998xn}, but here it emerges
as the only residual term if typical scalars acquire vevs as in
\eqref{vev} and the moduli and dilaton are stabilized. Taking $l_i =
l_0$ as a crude approximation, we find
\bea
V^{\frac{1}{4}} &\approx& |u_+|^\l\left(
\frac{le^g}{1+b_{E_8}l}\right)^{\frac{1}{4}}_{l_0}
\nonumber \\
&\approx& |u_+|^{\l + \frac{1}{2}}
A^{-\frac{1}{2}}\exp\left[\frac{1}{8\pi \a b_+}\right]_{l_0}
\label{scale}
\eea 
where in the second line I have used eq. \eqref{cond} and neglected
${\mathcal O}(1)$ factors from the $t_I$ moduli and Yukawa coupling
terms. This may be evaluated using eqs. \eqref{alpha}-\eqref{mgrav}
once the full group structure, matter representations and Yukawa 
couplings of the hidden sector are specified. However, it is clear that 
a very wide range of inflationary scales are possible using the general 
scheme \eqref{vev}. Purely as an example, a particularly natural choice 
might be to take $A \approx 1$, $b_+ \approx 0.05$ and $\l = 1/3$ so that 
$|\f_{AI}| \propto \L_c$. This gives $V \approx 10^{14}$ GeV, which is 
comfortably compatible with the COBE bound \eqref{COBE}.

%%%%%%%%%%%%%%%%%%%%%%%%%%%%%%%%%%%%%%%%%%%%%%%%%%%%%%%%%%%%%%%%%%%%%%%%%

\section{Conclusions}

In this paper I have argued that the predictions for the spectral
index of fluctuations $n_k$ that follow from inflationary models
usually assume that the GUT scale coupling was the same during
inflation as today. This could be false if the dilaton was stabilized
at a different value during inflation, and we could even lose all
calculability if either the theory became strongly coupled or there
was no (non-trivial) dilaton minimum. I have not addressed the
complicated issue of why we should expect the theory to have a
weak-coupling description in the first place~\cite{Dine:2000ds}, but
merely assumed that it does. I have then shown that, if dilaton
stabilization is due to corrections to the K\"ahler
potential~\cite{Banks:1994sg,Binetruy:1996nx}, then during inflation
the dilaton will be stabilized at a value which causes negligible
change in $\a$, provided the inflationary energy density arises from
vevs of the form \eqref{vev}. I do not claim that this is the only way
of achieving such a result. Indeed
in~\cite{Gaillard:1998xx,Cai:1999aj} other mechanisms were
proposed. Nor do I claim that it is not possible to achieve such
results using racetrack models~\cite{deCarlos:1992pd} although perhaps
the ``competing condensates'' makes  it less likely. However, the
scheme presented here is both natural and generic, relying on the same
mechanism as is known to provide vevs in the visible
sector. Additionally, I have not assumed any particular form for the
non-perturbative correction $g(l)$, except to suppose that it can
provide reasonable phenomenology today. Therefore it is perhaps
worthwhile to see if we can understand these results on a slightly deeper 
level.

In the absence of any corrections to the K\"ahler potential, we would
find $\a$ rising as $l$, but \eqref{da} shows that corrections of the form
\eqref{coupling} lead to a flattening of this dependence, which is
tuned to occur around the phenomenologically acceptable value. Of
course, as $l$ departs from $l_0$, higher derivative terms in
\eqref{da} will eventually cause significant changes in $\a$. The
presence of the ``flattened slope'' in $\a$ is precisely a consequence
of the vanishing of the cosmological constant in today's vacuum, as
expressed in \eqref{V=0}-\eqref{Vmin}. Why is this so? The question
essentially asks why the (ln of) the coupling should be the K\"ahler
potential for the dilaton. This is of course just a consequence of the
supersymmetry structure of the higher dimensional supergravity theory;
$D=10$ SUGRA requires a dilaton to complete its supermultiplet
structure, and supersymmetry then dictates how its K\"ahler potential
and coupling to other fields are related, whatever frame is used. Now,
the condensate itself depends on the dilaton only through the K\"ahler
potential and $\a$ as in \eqref{cond}. Of course, this must be
so. Therefore it is clear that there will be a flattening of the
condensate's dilaton dependence near today's value $l_0$ (this is
precisely what was tuned in \eqref{V=0}-\eqref{Vmin}). Hence, if we
can find a way to make the inflationary energy density depend on the
condensate in roughly the same way as the residual form of the scalar
potential today, we should expect it to have a flattened dependence
near $l=l_0$. Such is the idea behind \eqref{vev} and it is pleasing
that this is a very natural expectation. Can we make the inflationary
dilaton minimum $l_i$ close enough to $l_0$ that $\a$ is unaffected?
\eqref{lgprime} shows that the answer is yes, because the same higher
order (logarithmic) derivatives that destabilize $\a$ also describe
the difference between the $l_i$ and $l_0$. Clearly, this is because
the corrections to $\a$ and the K\"ahler potential are related as in
\eqref{Einst} which is simply a consequence of supersymmetry. Now,
both sides of \eqref{lgprime} are $\ll 1$ and so we need to ``mix in''
far less of the higher order derivative terms in order to satisfy this
compared to the amount that would cause an appreciable variation in
$\a$.

Although this paper is written with hybrid inflation in mind, such as
commonly arises in string inspired constructions, no attempt has been
made to specify any inflationary model. In particular, I have not
identified a slow-roll field, or the slope of the potential down which
it evolves and this deserves some comment. Candidate slow-roll fields
could arise from a number of sources. Firstly, it is possible to
imagine that whatever set up the condition \eqref{vev} is not
completely stable, so the inflaton is some linear combination of the
matter scalar fields. Secondly, we could assume that a vev of the form
\eqref{vev} is not in fact generic, with only certain fields taking
this form, others being driven to zero. In this case, one could find a
situation where one (combination) of the $t_I$ moduli is left
undetermined by the dominant terms in the potential (see {\it e.g.}
\cite{Copeland:1994vg}) thus providing a slow-roll field. These issues
have not been dealt with here. However, we have shown that with
\eqref{vev} the dilaton and moduli are minimized so as to provide a
constant energy density. Whatever terms destabilize this and lead to
the true vacuum must arise from terms in \eqref{V} shown to be
``negligible'' in comparison to those kept in \eqref{Vinfl2}. Hence,
they are bound to lead to {\it slow}-roll, at least in the vicinity of
the potential \eqref{Vinfl2}. The assumption here is that the vevs
\eqref{vev} can be preserved in that form on a timescale comparable
to, or larger than, the slow-roll timescale. The general theme here
was not to build new inflationary models, but rather to try to provide
some justification for the predictions of standard
ones~\cite{Lyth:1998xn}, should they be implemented in a string
context. With K\"ahler stabilization of the dilaton, this appears to
be quite plausible. It would be interesting to know whether racetrack
models, or more exotic ideas, lead to the same results. In particular,
it would be of great interest to know whether this can arise models of
gaugino condensation in the strongly-coupled heterotic
M-theory~\cite{Lukas:1997rb} where the necessity of stabilizing the
dilaton and moduli together becomes more apparent. Finally, it is
clearly important to try to pursue these ideas into post-inflationary 
epochs so as to attempt to make contact with apparent experimental 
results~\cite{Webb:2000mn,Banks:2001qc,Bahcall:2003rh}.

%%%%%%%%%%%%%%%%%%%%%%%%%%%%%%%%%%%%%%%%%%%%%%%%%%%%%%%%%%%%%%%%%%%%%%%%%

\section{Acknowledgments}

I would like to thank Subir Sarkar, John March--Russell and Oscar
Vives for very helpful conversations.

%%%%%%%%%%%%%%%%%%%%%%%%%%%%%%%%%%%%%%%%%%%%%%%%%%%%%%%%%%%%%%%%%%%%%%%%%%

%%%%%%%%%%%%%%%%%%%%%%%%%%%%%%%%%%%%%%%%%%%%%%%%%%%%%%%%%%%%%%%%%%%%%%%%%%

\end{document}